\documentstyle[prl,twocolumn,aps]{revtex}
\draft
\begin{document}
\title{Time correlations in 1D quantum impurity problems.}
\author{F. Lesage, H. Saleur$^*$ and  S. Skorik}
\address{Department of physics, University of Southern California,
Los-Angeles, CA 90089-0484.}
\date{November 28 1995.}
\maketitle

\begin{abstract}
We develop in this letter
an analytical approach using  form-factors to compute   time dependent correlations
in integrable quantum impurity problems.
As an example, we obtain for the first time   the frequency dependent
conductivity $G(\omega)$ for the tunneling between edges in the
$\nu=1/3$ fractional quantum Hall effect, and the spectrum $S(\omega)$
of the spin-spin correlation  in the  the anisotropic
Kondo model
and equivalently in the double well  system  of  dissipative quantum mechanics, both at vanishing temperature. 

\end{abstract}
\vskip 0.2cm

\narrowtext

Several problems of  great physical interest
can be reformulated as impurity problems in one dimensional
Luttinger liquids. These include the anisotropic Kondo problem, the double
and multi-well problem in dissipative quantum mechanics, the tunneling
between edges in the fractional quantum Hall effect. Although these
problems are integrable, exact  results have for a long time
mostly  concerned
static,  thermodynamic quantities. A recent approach
using massless scattering has also allowed the computation
of DC transport properties and noise,  in and out of equilibrium.
Unfortunately, AC properties had so far
remained inaccessible, while containing most of the
open physical questions. In this letter, we develop
an approach based on ``massless form-factors''  to compute the AC properties.
We illustrate it on two simple cases, both at vanishing
temperature: the  frequency dependent
conductivity $G(\omega)$ for the tunneling between edges in the
$\nu=1/3$ fractional quantum Hall effect, and the spectrum $S(\omega)$
of the spin-spin correlation  in the
double well problem of  dissipative quantum mechanics and in the anisotropic
Kondo model.

Edge excitations in the fractional
 quantum Hall effect  with filling fraction
$\nu=1/(2p+1)$ are thought to be described by a chiral Luttinger liquid
\cite{wen}.
This  allows a number
of interesting theoretical predictions
to be made, in particular
concerning the effect of impurities
\cite{kanefish}. For a single impurity (as can be obtained
 experimentally through
a constriction), one expects,
based on perturbation theory and scaling arguments,
 the (dimensionless)
DC conductance to behave as $c \ T^{2(1/\nu-1)}$ at low temperature
and
$\nu+c' \ T^{2(\nu-1)}$ at high temperature.
The problem
 can be described
by the hamiltonian~:
\begin{eqnarray}
H=\frac{1}{8\pi \nu}\int_{-\infty}^\infty dx \ [\Pi^2+
&(\partial_x \phi)^2]_R+[\Pi^2+
(\partial_x \phi)^2]_L\nonumber \\
&+\lambda  \ \cos[\phi_L(0)-\phi_R(0)]
\end{eqnarray}
where $\phi_{L/R}(x\pm t)$ are left and right moving field
moving on different edges.  This model can be
simplified  through the change of variables~:
\begin{eqnarray}
\phi^e(x,t)=\frac{1}{\sqrt{2}} [\phi_L(x,t)+\phi_R(-x,t)] \nonumber
\\
\phi^o(x,t)=\frac{1}{\sqrt{2}} [\phi_L(x,t)-\phi_R(-x,t)].
\end{eqnarray}
The even field factorises and after folding the real line we
get a hamiltonian on the half line with interaction at the
boundary which can be rearranged in the form~:
\begin{equation}
H=\frac{1}{16 \pi \nu} \int_{-\infty}^0 dx \
[\Pi^2+(\partial_x \phi)^2]+
\lambda  \ \cos[\phi(0)/2].\label{bdrSG}
\end{equation}
This hamiltonian is integrable, and the problem is thus
``solvable'' in principle. The DC
conductance \cite{fls} and  the  DC shot noise
\cite{flsn},\cite{fs} have been computed in the
general case of a non vanishing voltage. These
results followed from combining the standard thermodynamic Bethe ansatz
with a Boltzmann equation.

The problem of a two state system in a dissipative bath can
also be mapped onto a boundary problem \cite{sudrev}.
When the dissipation
is ohmic, it can described by a single spin interacting with
a bath of electrons.
The corresponding hamiltonian is the anisotropic Kondo
model, which is also known to be integrable:
\begin{eqnarray}
H=\frac{1}{16\pi \alpha} \int_{-\infty}^0 dx  [\Pi^2+(\partial_x\phi)^2]
&\nonumber\\ 
+ \lambda \ ( \sigma_+ e^{i\phi(0)/2}+&\sigma_- e^{-i\phi(0)/2}
),\label{anisK}
\end{eqnarray}
where $\sigma_\pm$ are Pauli matrices. The models (\ref{bdrSG}) and (\ref{anisK}) have the same ``bulk'' part:
a free boson (we denote  the coupling $\nu$ or $\alpha$ by $g$ in the
general discussion to follow), but  they differ significantly  by the boundary
interaction. They both  belong to the same large class of integrable boundary
field theories, and can be approached by a unified formalism \cite{flsrecent}.
The general strategy  is to describe the bulk part
by  a basis of states that scatter in a very simple way on the boundary. To
find the right basis,
one can think of more general hamiltonians that look like
(\ref{bdrSG}) and (\ref{anisK}) but with an additional term of the form
$\Lambda \int_{-\infty}^0 dx 
\cos[\phi(x)]$. The bulk part is then the well known sine-Gordon model, while
it can be shown that
the boundary interaction does not spoil integrability \cite{ghozamo}. The
right basis for the sine-Gordon model is well known, it consists of 
(massive) solitons/antisolitons, 
and for $g\leq 1/3$, $[1/g-2]$  breathers.  These have
factorized
scattering, and they  also scatter simply, without particle production, at the
boundary. One can now take the limit $\Lambda\to 0$ to obtain
a description of the free boson using massless solitons/antisolitons
and breathers \cite{F}.  This is of course
a more complicated basis than the standard plane waves, but
in the latter,  the
effect of the boundary interaction  is essentially intractable, while
here it is transparent. Massless particles have dispersion relation $E=p$
(resp. $E=-p$) for
right (R) (resp. left (L) ) moving particles. Since the theory is massless, we
can set the
arbitrary energy scale equal to unity, and parametrize the energies by a
rapidity: $E=e^\beta$ for
solitons/antisolitons, $E=2\sin( n\pi g  e^\beta/2(1-g))$ for the $n^{th}$
breather.

In what follows, we work in Euclidian space and choose $x$
to be the imaginary time. This is a ``modular transformed point of view'',
where the
boundary interaction does not appear in the hamiltonian any more,
but is encoded in  a
boundary state $|B>$, so correlators
can be represented as $<0|{\cal O}{\cal O}|B>$.
The boundary state has a simple expression  \cite{ghozamo} in terms of
solitons/anti-solitons and breathers creation operators,
$Z^*_\epsilon(\beta)$, with $\epsilon=\pm,1,...,[1/g-2]$~:
\begin{eqnarray}
|B>=\sum_{n,\epsilon's}\int
K^{\epsilon_1\epsilon_1'}(\beta_B-\beta_1)\ldots
K^{\epsilon_n\epsilon_n'}(\beta_B-\beta_n)
\nonumber \\
Z^{*}_{\epsilon_1 L}(\beta_1)\ldots
Z^{*}_{\epsilon_nL}(\beta_n)
Z^{*}_{\epsilon_1'R}(\beta_1)\ldots
Z^{*}_{\epsilon_n'R}(\beta_n)|0>,\label{bdstate}
\end{eqnarray}
where the integrals run on $[-\infty,\infty]$ and
are ordered $\beta_1<\ldots<\beta_n$. The boundary interaction is completely
encoded in the  matrix $K^{ab}$, which derives from the
reflection matrix, solution of the boundary Yang-Baxter equation:
\begin{equation}
K^{ab}(\beta)=R^a_{\bar{b}}\left(\frac{i\pi}{2}-\beta\right).
\end{equation}
The strength of the boundary interaction is encoded
in an energy scale $T_B=e^{\beta_B}$. The latter
is related in a non-universal way to the microscopic coupling, $T_B\propto
\lambda^{1/(1-g)}$.

To compute a correlation function in this approach, one needs to know the matrix
elements of the operators in the massless particle
basis, the so-called form factors \cite{smirnov}.
Away from $g=1/2$, there are infinitely such
 matrix elements to compute, since the theory is truly interactive. For
instance,
the $U(1)$ current acting on the vacuum can create
arbitrary numbers of solitons/antisolitons pairs and  breathers. There are thus
two difficulties:
to compute the matrix elements, and to sum their contributions.
In this letter, we will show how to compute the form-factors
by taking appropriate limits of the massive sine-Gordon form-factors. We will
also show that expansions in  multiparticle processes converge
extremely fast, so in practice only a few terms are necessary to
obtain excellent precision ($10^{-3}$) on the quantities of interest,
and this all the way from the short distance to the large distance fixed point.

We start with the frequency dependent conductance
in the problem of tunneling between edge states.
Using   Kubo formula  \cite{kanefish} and folding the system as
explained above one
finds $G={g\over 2}+\Delta G(\omega)$, where~:
\begin{eqnarray}
\Delta G(w_M)=\frac{1}{8 \pi \omega_M L^2}\int_{-L}^0 dx dx'
\int_{-\infty}^\infty d y \ e^{i \omega_M y}\nonumber\\
 \left[<\partial_z\phi(x,y)\partial_{\bar{z}'}
\phi(x',0)> +cc\right],\label{matsub}
\end{eqnarray}
$\omega_M$ being a Matsubara frequency,  $z=x+iy$. We have used here the fact
that
 in the current-current correlation
the term $<\partial_z\phi\partial_{z'}\phi>$
and its conjugate are insensitive to the impurity \cite{affwong}, contributing
${g\over 2}$ to the
conductance. This is because $\partial_z\phi$ (resp. $\partial_{\bar{z}}\phi$)
act only on R (resp. L) particles, while the  effect of the  boundary is to mix
L and R-particles. Considering a particular term in the expansion of $|B>$ with
$n$ L- and $n$ R-particles, the only process with a non vanishing amplitude in
$<0|\partial_z\phi\partial_{\bar{z}'}\phi|B>$ is to have $\partial_{\bar{z}'}\phi$
annihilate
all the $n$ L-particles and then $\partial_z\phi$ all the $n$ remaining
R-particles.
We thus need the form-factors~:
\begin{eqnarray}
f(\beta_1,...,\beta_n)_{\epsilon_1,...,\epsilon_n}=& \nonumber \\
<0|\partial_z \phi(0) \ Z^{*}_{\epsilon_1R}&(\beta_1)\ldots
Z^{*}_{\epsilon_nR}(\beta_n)|0>.\label{tata}
\end{eqnarray}
Current form-factors have been
computed by Smirnov \cite{smirnov} for the massive sine-Gordon
model. Expressions for (\ref{tata}) can then be obtained by taking a   massless
limit, ie by
sending the
physical mass to zero and the rapidities to infinity to
keep excitations of  finite energy \cite{mussardo}.

As examples, when $g=1/2$, there is only a pair of solitons/antisolitons
in the spectrum.  The only non zero form factor is
$f_{\pm\mp}(\beta_1,\beta_2)=\mp c \ e^{(\beta_1+\beta_2)/2}$, with
$c$ a known normalisation.
For $g=1/3$, a breather appears in the spectrum and the
first two form factors
are given by~:
\begin{equation}
f_1(\beta)=c_1 e^\beta
\end{equation}
for the 1 breather and~:
\begin{equation}
f_{\pm,\mp}(\beta_1,\beta_2)=\mp c_2
\frac{\zeta(\beta_1-\beta_2)}
{\cosh(\beta_1-\beta_2)} \ e^{(\beta_1+\beta_2)/2},
\end{equation}
for the two solitons form factors.
Here $\zeta(\beta)$ is a known
function  and $c_1,c_2$ known constants, whose expressions we will give
elsewhere \cite{a-venir}.

{}From (\ref{bdstate}), the form factors expansion results
into the  general expression
\begin{eqnarray}
<\partial_{\bar{z}}\phi(x,y)\partial_{z'}\phi(x',y')>=&\nonumber
\\
\int_0^\infty dE \ {\cal G}(E) \exp[E&(x+x')-iE(y-
y')],\label{fourep}
\end{eqnarray}
and from (\ref{matsub})~:
\begin{equation}
\Delta G(\omega)=\frac{1}{4\omega} {\cal I}m {\cal G}(-i\omega).
\end{equation}

Using the previous formulaes and the expression for the
form factors and the boundary scattering matrices\cite{ghozamo},
we find~:
\begin{equation}
G(\omega)=\frac{1}{2}\left[1-(T_B/\omega){\rm tan}^{-1}(\omega/T_B)\right],
\end{equation}
for $g=1/2$ in agreement with  previous results \cite{guinea}.
For $g=1/3$, there is an infinity of form factors contributing
to the correlation. However,
when computing the conductivity,
we find a very rapid convergence with the number of rapidities;  two rapidities
are
sufficient to get a $1\%$ accuracy. Let us stress that
this convergence is independent of the strength of
the impurity, and the results are valid for the {\bf whole} flow
from small to large distances.  The quantity $G(\omega)$ is
plotted in figure 1 as a function of $T_B/\omega$.

Some general features of $G(\omega)$ can easily be deduced from this approach.
The  reflection
matrices of solitons/antisolitons expand as a double power series in
$\exp\beta$ and
$\exp({1\over g} -1)\beta$, the reflection matrix of breathers
as a power series in $\exp\beta$. This leads to a double power series in
$(\omega/T_B)^{-2+2/g}$
 and $(\omega/T_B)^2$ at small frequencies, $(T_B/\omega)^{2-2g}$ and
$(T_B/\omega)^2$ at large frequencies.
Therefore, as first argued in \cite{guinea2}, at low frequencies, $G(\omega)$
goes as $\omega^2$ for $g<1/2$
and $\omega^{-2+2/g}$ for $g>1/2$.

The same method can be applied to compute the spin spin correlation
$C(t)=\frac{1}{2} <[\sigma(t),\sigma(0)]>$
or its fourier transform conventionally denoted
$\chi''(\omega)$ in the two state problem \cite{sudrev}.
A difficulty arises at first sight because the massless scattering
description of the anisotropic Kondo problem
involves only the massless sine-Gordon particles and no spins (physically, this
is
because this description is based on the large distance limit of the theory
where the spin is screened).
However, using the fact that spin-flips, which are induced by $\sigma_\pm$, are
coupled
with insertions of vertex operators at the boundary, one
can relate $C(t)$ to the current current correlator, and get the expression~:
\begin{equation}
\chi''(\omega)={1\over (2g\pi)^2}{1\over\omega^2}
{\cal I}m\left[{\cal G}
(-i\omega,T_B)-
{\cal G}(-i\omega,0)\right],
\end{equation}
where ${\cal G}$ is the Fourier transform of the L-R current correlator
defined in (\ref{fourep}).
The only difference with the conductance problem  is in the boundary
interaction.  The reflection matrix
for the solitons are now given by \cite{F}~:
\begin{equation}
R_\pm^\pm=\tanh\left(\frac{\beta}{2}-\frac{i\pi}{4}\right), \ \ R^\pm_\mp=0,
\end{equation}
and for the breathers we find \cite{a-venir}~:
\begin{equation}
R^m_m=\frac{\tanh(\frac{\beta}{2}-\frac{i\pi m}{4 (1/g-1)})}
{\tanh(\frac{\beta}{2}+\frac{i\pi m}{4 (1/g-1)})}.
\end{equation}

The computations are then done along the same
lines as  before.  For $g=1/2$, the free or
Toulouse point,
the previously known result~:
\begin{equation}
\chi''(\omega)
={1\over \pi^2}
{4T_B^2\over  \omega^2 +4T_B^2}\left[{1\over\omega}
\ln\left({T_B^2+\omega^2\over T_B^2}\right)+{1\over T_B}
\tan^{-1}{\omega\over T_B}\right]
\end{equation}
is recovered.
For a whole domain  of $g\in [.25, .6] $,  the
form factor expansion
gives again  very precise results  for $\chi''(\omega)$
with only the first two
terms, for all strengths of the impurity coupling .  Let us give as an example the explicit expressions for
$g={1\over 3}$. The contribution from the breather is~:
\begin{equation}
\frac{c}{\omega}{\cal R}e 
\left[R^1_1(\log(\frac{\omega}{\sqrt{2}T_B}))-1 \right]
\end{equation}
with $c=-0.141$ and the contribution for the two solitons form factors
is given by~:
\begin{eqnarray}
\frac{c'}{\omega}{\cal R}e \int_{-\infty}^0 d\beta
\frac{\vert \zeta(\beta-\log(1-e^\beta))\vert^2}{
\cosh^2(\beta-\log(1-e^\beta))} e^\beta &
 \\
\left\{  R^+_+(\beta+
\log(\frac{\omega}{T_B})) R^+_+(\log[(1-e^\beta)\frac{\omega}{T_B}])
-1\right\} ,\nonumber
\end{eqnarray}
with $c'=-0.0451$ and the function $\zeta(\beta)$ given in
\cite{smirnov}.  In figure 2 we show the function
$S(\omega)=\chi''(\omega)/\omega$ for $g=\frac{1}{4},
\frac{1}{3},\frac{1}{2},\frac{3}{5}$ at $T_B=0.1$.  From the
scaling form $S(\omega)=\frac{1}{\omega^2} F(\omega/T_B)$,
the results for all $T_B$ are recovered.  We observed with very
good precision that the ``quasi-particle peak'' \cite{sudrev} disappears at
$g={1\over 3}$ (instead of the sometimes conjectured $g={1\over 2}$),
but we have no analytical proof of this result.  Physically this
means that the crossover between the oscillating behaviour and 
the damped one is at $g={1\over 3}$ at zero temperature.

It is easy to check from the form factor expansion that for
all $g$ the following relation holds~:
\begin{equation}
\lim_{\omega\rightarrow 0} {\chi''(\omega)\over\omega}
= \frac{1}{\pi^2 T_B^2 g},
\end{equation}
which leads to the large time asymptotics
$C(t)\approx -{\sqrt{\pi}\over e g}{1\over (\pi T_B t)^2}$.

At small frequency, $S(\omega)$ expands as a power series in
$(\omega/T_B)^{2n}$ for any $g$.
At large frequency, it expands as a double series in $(T_B/\omega)^2$ and
$(T_B/\omega)^{2-2g}$. As
a consequence, $C(t)\propto t^{2-2g}$ at small times.

In conclusion, we have shown that the description of quantum impurity problems
based on $(i)$ massless integrable boundary field theories, $(ii)$ boundary
states, and $(iii)$ form-factors, allows an efficient computation of the
 time dependent properties. So far, these had been completely inaccessible
except
by Monte Carlo simulations, and by various approximations which were recently
proven unreliable \cite{sudjoe}. Our goal here was to present the technique and
some results.
Details will be provided in a more extensive paper. We also expect to be able
to
extend the method  to the case where there is a voltage  (a bias) and to the
finite temperature case. This will involve a combination of both the
form-factors
and the TBA approach. Also, the careful reader might wonder why the computation of AC properties requires form-factors, while DC properties have been successfully computed so far by analogy with the free case. The reason is that DC properties see only the charge $Q$, ie the $x$ integral of the current, and 
 $Q$ can be shown to act  purely diagonally on multiparticle states. See \cite{a-venir} for more details.

\vskip .2in
\noindent $^*$ Packard Fellow.
\vskip .2in
We thank P. Fendley, C. Mak and S. Chakravarty for many useful discussions.
This work was supported by the Packard Foundation, the
National Young Investigator program (NSF-PHY-9357207) and
the DOE (DE-FG03-84ER40168). F. Lesage is partly supported by
a Canadian NSERC Postdoctoral Fellowship.

\newpage 

\begin{figure}
\caption{Frequency dependent conductivity at $T=0$, $g=1/3$.}
\label{fig1}
\end{figure}

\begin{figure}
\caption{Response function for different values of $g$.}
\label{fig2}
\end{figure}

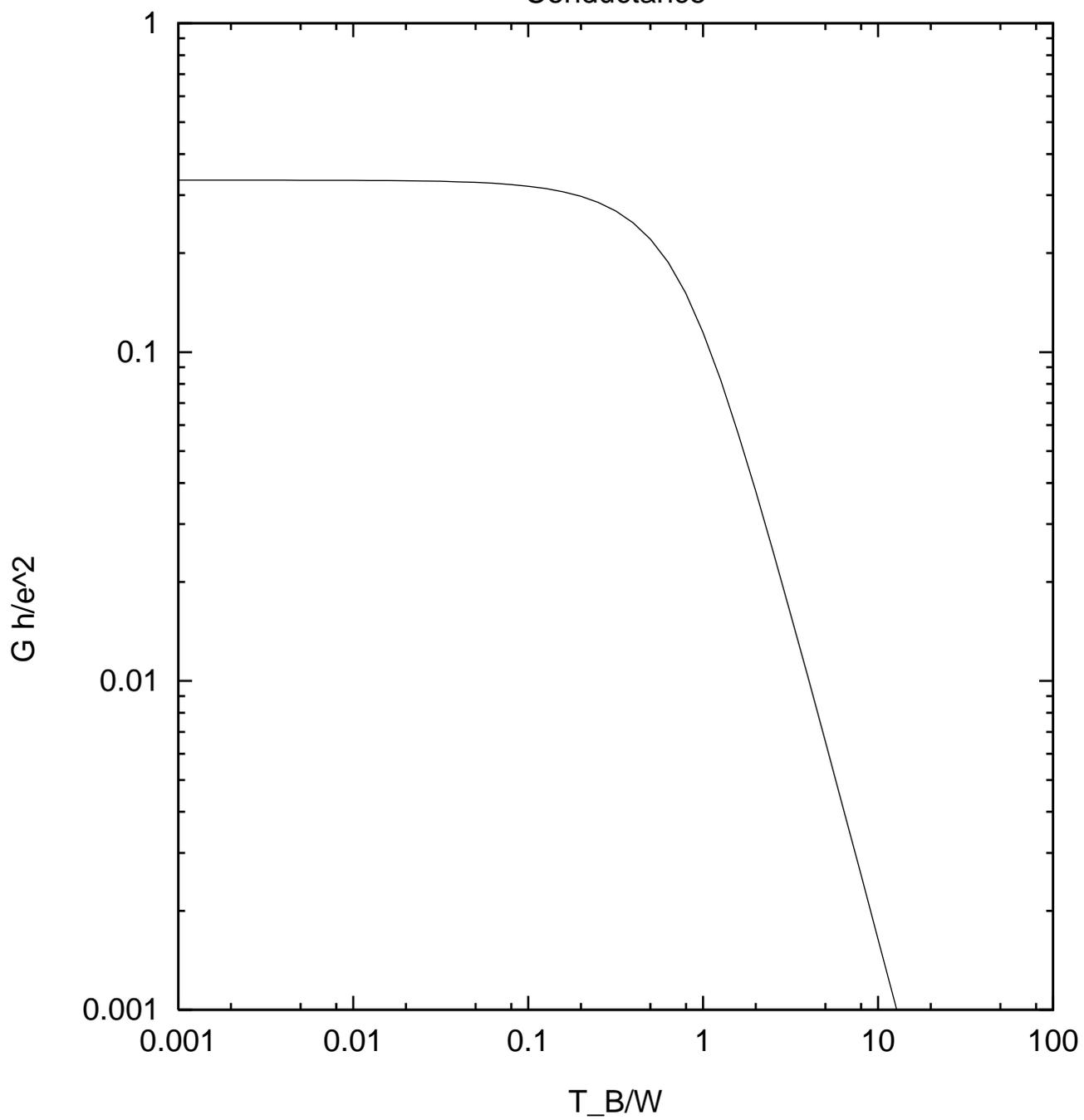
\end{document}